\documentclass[review,3p]{elsarticle}
\usepackage{hyperref}
\usepackage{caption}
\usepackage{subcaption}
\journal{Journal of Physics of the Dark Universe}

\newcommand{\hs}[1]{{\hat s_{#1}}}
\newcommand\beq{\begin{eqnarray}}
\newcommand\eeq{\end{eqnarray}}
\newcommand\bq{\begin{equation}}
\newcommand\eq{\end{equation}}

\DeclareSymbolFont{matha}{OML}{txmi}{m}{it}% txfonts
\DeclareMathSymbol{\vv}{\mathord}{matha}{29} 
\begin{document}

\begin{frontmatter}

\title{Polarized target as a tool for probing the non-standard properties of dark matter}
%	\tnoteref{mytitlenote}}
%\tnotetext[mytitlenote]{Fully documented templates are available in the elsarticle %package on \href{http://www.ctan.org/tex-archive/macros/latex/contrib/elsarticle}{CTAN}.}

\author{Arkadiusz B\l{}aut}
	\ead{ arkadiusz.blaut@uwr.edu.pl}
	
	\author{Wies\l{}aw Sobk\'ow\corref{mycorrespondingauthor}}
	\cortext[mycorrespondingauthor]{Corresponding author}
		\ead{ wieslaw.sobkow@uwr.edu.pl}
		
		\address{Institute of Theoretical Physics, University of Wroc\l{}aw,
				Pl. M. Born 9, PL-50-204~Wroc{\l}aw, Poland}

\begin{abstract}
Possibility of using the polarized  nucleus target for testing the non-standard properties of fermionic dark matter 
is considered. It is assumed that the incoming dark matter scatters on the polarized nuclei in the presence of vector, 
axial, scalar and tensor interactions. The scalar and tensor couplings are assumed to be complex numbers. 
Various types of measurement settings are tested depending on the assumptions regarding dark matter polarization 
and measurement of  the recoil nucleus spin. In particular we address the question of finding and distinguishing effects of theoretically possible scenarios in which mixed triple products among various  polarization vectors, the momenta of incoming DM and of recoil nucleus appear in the cross section. Their presence would  indicate the possibility of  time reversal symmetry violation in the polarized dark matter-nucleus scattering. Our considerations are model-independent and carried out in the non-relativistic dark matter and recoiled nucleus limit. 
\end{abstract}


\end{frontmatter}

%\linenumbers

\section{Introduction}
\label{sec1}

The astronomical and cosmological data indicate the presence of dark matter (DM), but this evidence is indirect and based on the gravity impact  on stars, galaxies, galaxy clusters, the cosmic microwave background radiation. The particle nature of DM is still not identified. Let us recall that the Standard Model(SM) \cite{SM,SM1,SM2,SM3,SM4} does not explain  the concepts of DM and dark energy, and  some fundamental issues such as:   the origin of  parity violation, the violation of  combined symmetry of charge conjugation and space inversion (CP), observed  baryon  asymmetry between matter and antimatter of universe \cite{barion} through a single CP-violating phase of the Cabibbo-Ko\-ba\-ya\-shi-Mas\-ka\-wa quark-mixing matrix (CKM) \cite{Kobayashi}. There is also a lack of clarity about the hierarchy problem (the enormous difference between the weak and Planck scales in the presence of the Higgs field) and the spontaneous symmetry breaking associated with the origin of mass. In such a situation it is natural to suppose that SM is the low-energy limit of a more fundamental theory. This has led to a number of proposals for non-standard models, but none of them have proved conclusive. There are many  candidates for DM particles predicted by various non-standard  schemes. One of the most intensive and constantly investigated ideas is the  paradigm  of weakly interacting massive particles (WIMPs)  with masses around   the weak scale \cite{DM1,DM2,DM3,DM4}. WIMPs can  be made of heavy fourth-generation neutrinos, supersymmetric particles, particles in models with universal extra dimensions, etc. These may be scalar or fermion particles (Dirac or Majorana type) \cite{DM5,DM6,DM7,DM8}. 
Despite the efforts, there is no direct or indirect evidence of the existence of the WIMPs to date, but  new measurement strategies have been proposed to detect DM in the laboratory, in accelerators and spaces.  In this situation it makes sense to examine   also other theoretically possible  scenarios, such as a light DM (LDM), with masses in the  keV to GEV range. Interesting studies of the use of LDM scattering on electrons can be found in  \cite{DM9,DM10,DM11}.\\ The main goal of terrestrial direct-detection experiments is the measurement of energy spectrum and the direction of the nuclear recoils (or recoil electrons in the case of LDM)  coming from the interactions  of DM with ordinary particles \cite{DM12,DM13,DM14,DM15,DM16}. It must be noted that in these experiments there is an irreducible background produced by neutrinos, which can mimic the right signal. The key issue of new experiments will be the possibility of distinguishing between events from the scattering of DM and events generated by neutrino scattering in  the presence of non-standard interactions \cite{NuB1,NuB2,NuB3,NuB4,NuB5,NuB6}. The approach shown in \cite{NuB7} is worth noting, as it assumes that exotic interactions can communicate with both DM and neutrinos. Furthermore, neutrino oscillation effects  should also be taken into account.  The above ideas involve the un-polarized detection targets. 

As an alternative to experiments with un-polarized targets, the idea of using polarized target (PT) to detect DM is worth to be considered. It should be  stressed  that  PT  has been proposed to probe the flavour composition of  (anti)neutrino beam \cite{PET1}, the neutrino magnetic moments \cite{PET2,PET3}, the time reversal symmetry violation and neutrino nature in the (semi)leptonic processes \cite{PET4,PET5,PET6}, and most recently also for searching for DM \cite{PDM1,PDM0,PDM2}.  A basic advantage  of using PT is  an opportunity of changing the rate of weak interaction by inverting the direction of magnetic field. This feature is very essential in the measurement  of low energy neutrinos because the background level would be precisely controlled in the  direct detection DM experiments \cite{Misiaszek}.  It is  important to mention the prototype  measurements confirming  the possibility of building  the polarized target crystal of $Gd_{2}SiO_{5}$ (GSO) doped with Cerium (GSO:Ce) \cite{INFN}. It  is also  noteworthy  that there are known the methods  of  producing  the polarized  gasses such as helium,  argon and xenon \cite{Gass1,Gass2}.

In this paper,  we focus on using a polarized target to test the non-standard properties of fermionic DM 
in the presence of vector, axial, scalar and tensor interactions. In particular, we think about the possibility of breaking symmetry with respect to time inversion for the scenarios with the complex scalar and tensor couplings of DM. It is worth noting here an interesting analysis of the possibility of P- and CP-violation in  un-polarized DM-nucleus scattering \cite{PCP}.  We show how the energy and angular (polar-azimuthal) distributions of recoil nuclei are sensitive to the non-vanishing mixed triple products among various  polarization vectors, the momenta of incoming DM and of recoil nucleus. In addition, we check whether these observables can be used to differentiate between the Dirac and the Majorana DM.

\section{Scattering of fermionic dark matter on polarized target}
\label{sec2}

\begin{figure}[!h]
	\begin{center}
    \includegraphics[width=0.65\linewidth]{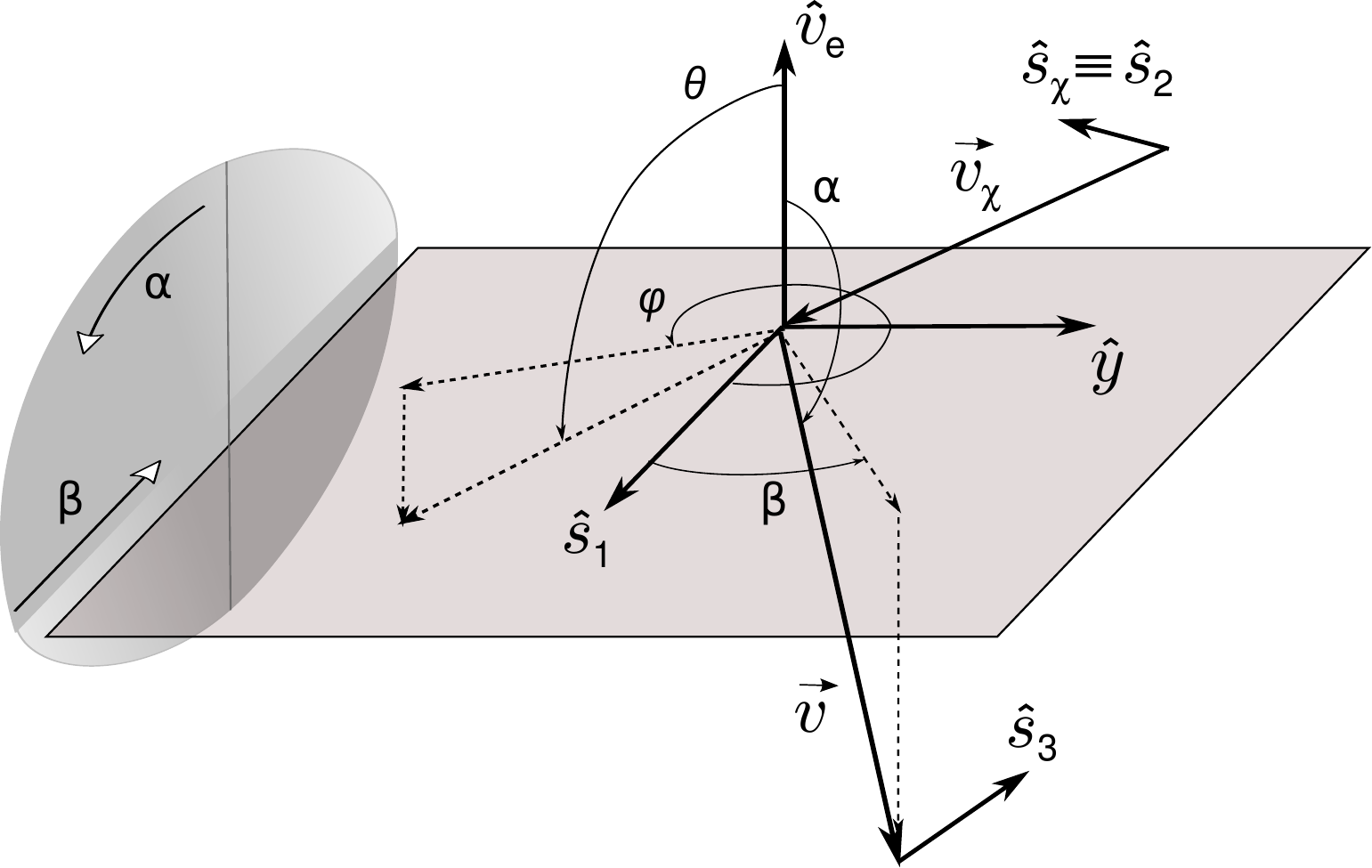}
	\end{center}
	\caption{Kinematics of polarized DM-nucleus scattering. $\hat{v}_{e}$ is  the Earth's velocity in the galactic rest frame; $ \hs{1}$ is the 	polarization vector of the  nuclear target;  $\vec{v}_{\chi}$ is the velocity of the incoming DM; $\hs{2}$ is the polarization vector of the incoming DM; $\vec{v}$ is the velocity of the recoil nucleus; $\hs{3}$ is   the 	polarization vector of the outgoing nucleus; $\alpha $, $\beta$ are the polar and azimuthal angles of  recoil nucleus velocity; $\theta$ and $\phi$ are the polar and azimuthal angles of   incoming DM velocity. The three vectors, $\left\{\hs{1},\hat{y},\hat{v}_e \right\}$, form an orthonormal frame.}
    \label{frame}
\end{figure}

We consider the  scattering of   fermionic  DM  of the Dirac type on the polarized nuclear target:   $\chi + N \rightarrow \chi + N$.  We assume that a DM fermion of mass $m_{\chi}$ interacts with a spin $1/2$  nucleus of mass $m$ via the four-fermion point interactions. The initial nucleus is polarized and  at rest.   For simplicity, the nucleus is treated  a point object (the nuclear effects are assumed to be contained in the effective couplings). However, it should be made clear here that a complete analysis of nuclear effects will be necessary for experiments with a specific target and may be the subject of a new study. 

 %We also assume  that the incoming DM %particle is the  superposition of left chiral  states %with right chiral ones. A direct %consequence of the superposition of the two chiralities %is the appearance  of the non-zero %transversal component of  DM  polarization vector. 
 
Fig. \ref{frame} illustrates the relative orientation of all vectors and explains  the notation for the relevant quantities. 
The amplitude for  this scattering is of  the form: 
\beq \label{ampDM}
M
&=& \left[\overline{u}_{\rm N'}\gamma^{\alpha}(c_{\rm V}
+ c_{\rm A}\gamma_{5})u_{\rm N}\right] \left[\overline{u}_{\chi'}
\gamma_{\alpha}(d_{\rm V} + d_{\rm A} \gamma_{5})u_{\chi}\right]\\
&  & \mbox{} +
\left[c_{\rm S_{\rm R}}\overline{u}_{\rm N'} ( 1+  \gamma_{5}) u_{\rm N} + c_{\rm S_{\rm L}}\overline{u}_{\rm N'} ( 1-  \gamma_{5}) u_{\rm N}\right]  \left[d_{\rm S_{\rm R}}\overline{u}_{\chi'}
( 1+  \gamma_{5})u_{\chi} +
 d_{\rm S_{\rm L}}\overline{u}_{\chi'}
(1 -  \gamma_{5})u_{\chi}\right]\nonumber\\
&& \mbox{} +
\frac{1}{2} \left[c_{\rm T_{\rm R}}
\overline{u}_{\rm N'}\sigma^{\alpha \beta} (1+\gamma_5) u_{\rm N} + c_{\rm T_{\rm L}}
\overline{u}_{\rm N'}\sigma^{\alpha \beta} (1-\gamma_5) u_{\rm N}\right]
\left[d_{\rm T_{\rm R}}(\overline{u}_{\chi'}\sigma_{\alpha \beta}(1
  + \gamma_{5})u_{\chi})
+ d_{T_{\rm L}}(\overline{u}_{\chi'}\sigma_{\alpha \beta}(1 - \gamma_{5})u_{\chi})\right] \nonumber.
\eeq
The coupling constants  $c_{\rm V} $,
$c_{\rm A}$, $d_{\rm V} $,
$d_{\rm A}$ corresponding to the vector and axial interactions are real numbers, while  the  non-standard 
$c_{\rm S_{\rm R,L}}, c_{\rm T_{\rm R,L}},  d_{\rm S_{\rm R,L}},  d_{\rm T_{\rm R,L}}$   couplings,  associated to  the scalar and tensor interactions, are the complex numbers denoted as
$c_{\rm S_{\rm R}} = |c_{\rm S_{\rm R}}|e^{i\,\phi_{\rm S_{\rm R}}}, d_{\rm S_{\rm R}} = 
|d_{\rm S_{\rm R}}|e^{i\,\theta_{\rm S_{\rm R}}}$, etc. 
Reality of  vector and axial coupling  constants follows from the  hermiticity of the interaction lagrangian. For the same reason, we also take into account the relations between the non-standard complex couplings with left- and right-handed chirality,
% appearing at the level of interaction lagrangian:
$d_{\rm S_{\rm L}, \rm T_{\rm L}}=(d_{\rm S_{\rm R}, \rm T_{\rm R}})^*$, $c_{\rm S_{\rm L}, \rm T_{\rm L}}=(c_{\rm S_{\rm R}, \rm T_{\rm R}})^*$.  
\par Our main goal to show how the presence of  non-standard couplings manifest themselves
in the angular and energy  distributions  of the recoil nucleus. This is mainly about the effects of a time reversal violation in  the polarized DM-nucleus scattering. To this end, we calculate the differential cross section, the differential event rate per unit detector mass and define the function  that purely depends  on the target polarization: 
\beq
	d\sigma &=&
	\frac{|M|^2}{64\,\pi^2\,m_{\chi}^{2}\,m\,v_{\chi}\,|\hat{v}_{\chi}\cdot\hat{v}|}\,\delta\left(v_{\chi}-\bar{v}
	\right)\,dE_{\rm R}\,d\beta\,d\cos{\alpha}, \\
%	&\Rightarrow \\
	\frac{d^2R}{dE_{\rm R}\,d\Omega}&=&
	\frac{\rho_{\chi}}{m_{\chi}\,m}\int d^3 v_{\chi}\,\frac{d^2\sigma}{dE_{\rm R}\,d\Omega}\,v_{\chi}\,f(\vec{v}_{\chi})  \nonumber \\
	&=& \sum\limits_{l=1,2}
	\frac{\rho_{\chi}}{64\,\pi^2\,m_{\chi}^{3}\,m^2}\int d\varphi\int d\cos{\theta}\,\,
	\frac{\bar{v}^2|\,M(E_{\rm R},\alpha,\beta,\theta,\varphi)|^2}{|\hat{ v}_{\chi}\cdot\hat{v}|}\,\Theta_{l}\,f(\vec{v}_{\chi}=\bar{v}\,\hat{v}_{\chi}), \label{r3}\\
%	&=& \qquad {\rm (2.8)\quad Catena}
%\end{eqnarray}
\label{r4}
%\frac{d^{2} R}{d E_{R} d \Omega} &=& \frac{\rho_{\chi}}{2\pi m_\chi m_{N}} \int %\frac{d\sigma}{d E_R}\delta(\vec{v}\cdot \hat{q}-v_{q}) v^2f(\vec{v})d^{3}v \\
%\frac{d\sigma}{d E_R}&=& \frac{1}{32\pi m_N m_{\chi}^2 v^2}  |\overline{M}|^2\\
\frac{d^{2}\Delta R}{d E_{\rm R} d\,\Omega} & = & \frac{1}{2}\bigg[\frac{d^{2} R(\hs{1})}{d E_{\rm R}\,d\Omega}  - 
\frac{d^{2} R(-\hs{1})}{d E_{\rm R}\,d\Omega}\bigg]. 
\eeq
$\rho_{\chi}\simeq 0.4\,$GeV$\,$cm${}^{-3}$ is the local DM density; 
\bq
f(\vec{v}_{\chi})=  \frac{1}{N} e^{-(\vec{v}_{\chi}+\vec{v}_e)^2/v_{0}^{2}}
\eq
(for $|\vec{v}_{\chi} + \vec{v}_e|\leq v_{\rm esc}$) is the DM velocity distribution in the laboratory frame (Maxwell-Boltzmann distribution truncated at the galactic escape velocity $v_{\rm esc}=544\,$km s${}^{-1}$);  $\vec{v}_e=232$\,km s${}^{-1}$ is the Earth's velocity in the galactic rest frame and is assumed to be constant magnitude; $N$ 
is the normalized factor, and $\Theta_{1}$, $\Theta_{2}$ are kinematical factors for the DM distribution, see \cite{PDM0};   $\mu$ is the  reduced mass of the DM-nucleus system; $q=m v$ is the nuclear recoil momentum; 
%$v_{q}= q/2\mu$ is the nuclear recoil velocity; 
$E_{\rm R}=q^2/2m$ is the nuclear recoil energy; $\bar{v}=q/[2\mu(\hat{v}_{\chi}\cdot\hat{v})] $; 
$d \Omega=d\cos\alpha\, d\beta$ is  the  infinitesimal
solid angle around the nuclear recoil direction;  $|M|^2$ is the squared modulus of the scattering amplitude. 
All results for the cross sections are expressed in the terms of right  chiral couplings for the scalar and tensor interactions of DM. The pseudoscalar interaction is omitted because it does not have a qualitative effect on results and conclusions. 
% \par Some clarification is needed for the integration of the  angular (polar-azimuthal) %distributions  over the azimuthal  $\phi$ and polar  $\theta$ angles of the momentum of the %incoming DM.  
 %Details of integration over the incoming DM velocities can be found in []. 
In addition, spectral functions  $(1/R) (dR/d E_{\rm R})$ and  $(1/|\Delta R|) (d|\Delta R|/d E_{\rm R})$ are used. 
All plots are made for the following choice of mass values and coupling constants: 
$m=m_{\chi}=\,$100 GeV,  $c_{\rm V} = -c_{\rm A}=d_{\rm V}= -d_{\rm A}=1/2$, 
$|c_{S_{\rm R}}|=|c_{\rm T_{\rm R}}|=|d_{\rm S_{\rm R}}|=|d_{\rm T_{\rm R}}|=1/2$; the overall factor standing at the detection rate, $\rho_{\chi}/(64\pi^2 N m_{\chi}^{3}m^2)$, is put to $1$.
 
 \section{First scenario - polarized target}
\label{sec3}

In this section we inquire how the presence of the vector, axial, tensor and scalar interactions in the polarized DM-nucleus scattering affects the spectrum and angular (polar and azimuthal) distributions of nuclear recoils and whether it allows for the appearance of terms that break the time reversal symmetry.

\begin{figure}[!h]
	\begin{center}
		\includegraphics[width=0.75\linewidth]{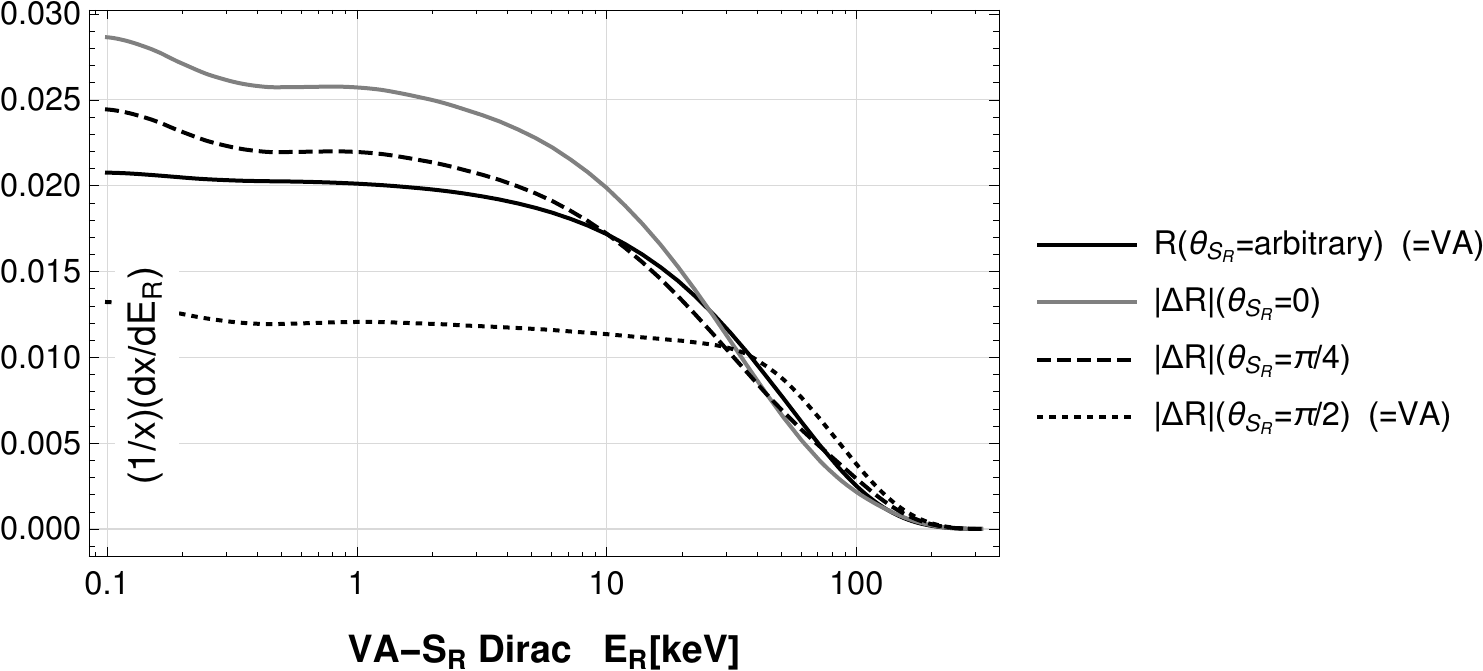}
	\end{center}
	\caption{Dirac DM: plot of  $(1/R) (dR/d E_{\rm R})$ as a function of $ E_{\rm R}$ for vector V, axial A and scalar S interactions with $\theta_{\rm S_{\rm R}}=\mbox{arbitrary}$  (solid black line); plots of $(1/|\Delta R|) (d|\Delta R|/d E_{\rm R})$ as a function of $ E_{\rm R}$   for V, A, S interactions; solid gray line for $\theta_{\rm S_{\rm R}}=0$; dashed line for $\theta_{\rm S_{\rm R}}=\pi/4$; dotted line for $\theta_{\rm S_{\rm R}}=\pi/2$.  }
	\label{E_VASR_Dirac}
\end{figure}

\begin{figure}[!h]
	\begin{center}
		\includegraphics[width=0.68\linewidth]{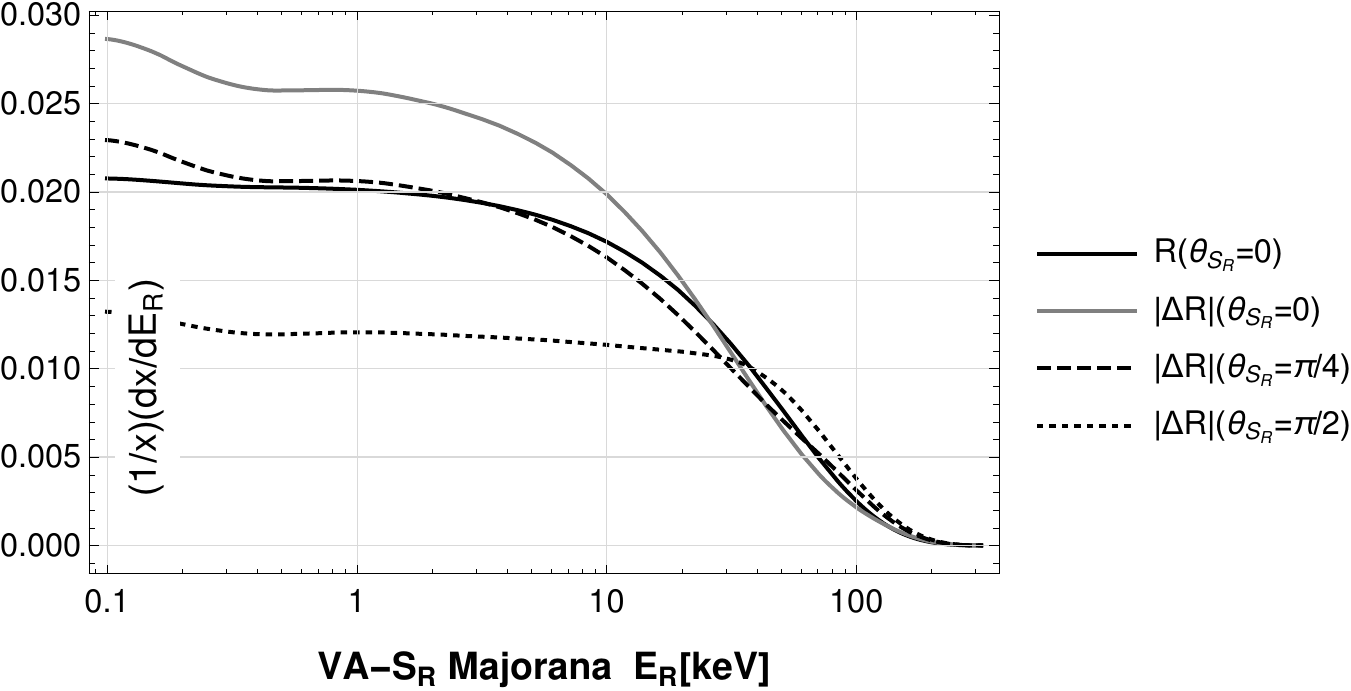}
	\end{center}
	\caption{Majorana DM: plot of  $(1/R) (dR/d E_{\rm R})$ as a function of $ E_{\rm R}$ for vector V, axial A and scalar S interactions with $\theta_{\rm S_{\rm R}}=0$  (solid black line); plots of $(1/|\Delta R|) (d|\Delta R|/d E_{\rm R})$ as a function of $ E_{\rm R}$   for V, A, S interactions; solid gray line for $\theta_{\rm S_{\rm R}}=0$; dashed line for $\theta_{\rm S_{\rm R}}=\pi/4$; dotted line for $\theta_{\rm S_{\rm R}}=\pi/2$. }
	\label{E_VASR_Majorana}
\end{figure}

\begin{figure}[!h]
	\begin{center}
		\includegraphics[width=0.68\linewidth]{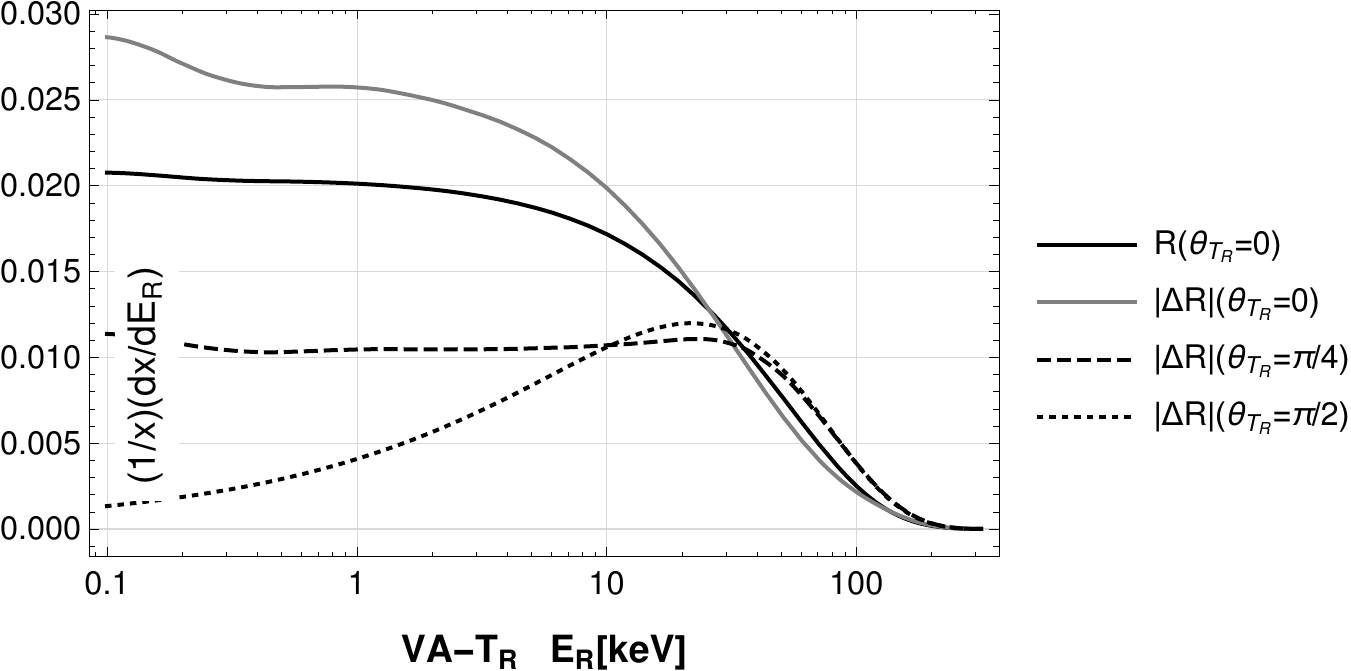}
	\end{center}
	\caption{Dirac DM: plot of  $(1/R) (dR/d E_{\rm R})$ as a function of $ E_{\rm R}$ for vector V, axial A and tensor T interactions with $\theta_{\rm S_{\rm R}}=0$  (solid black line); plots of $(1/|\Delta R|) (d|\Delta R|/d E_{\rm R})$ as a function of $ E_{\rm R}$   for V, A, T interactions; solid gray line for $\theta_{\rm S_{\rm R}}=0$; dashed line for $\theta_{\rm S_{\rm R}}=\pi/4$; dotted line for $\theta_{\rm S_{\rm R}}=\pi/2$. }
	\label{E_VATR_Dirac}
\end{figure}

In what follows we use the squared  amplitudes $|M_{\rm VA}(\hs{1})|^2$, $|M_{\rm VA-{S_R}}(\hs{1})|^2$, 
$|M_{\rm VA-{T_R}}(\hs{1})|^2$,
where 
\begin{eqnarray}
|M_{\rm VA-{S_R}}(\hs{1})|^2 &=& |M_{\rm VA}(\hs{1})|^2 + |M_{\rm S_R}(\hs{1})|^2 + 2\Re[M_{\rm VA-S_R}(\hs{1})] \\
|M_{\rm VA-{T_R}}(\hs{1})|^2 &=& |M_{\rm VA}(\hs{1})|^2 + |M_{\rm T_R}(\hs{1})|^2 + 2\Re[M_{\rm VA-T_R}(\hs{1})],
\end{eqnarray}
and difference of  squared  amplitudes (for all interactions)
\begin{equation}
 \Delta M(\hs{1})\equiv \frac12\left[|M(\hs{1})|^2- |M(-\hs{1})|^2\right].
\end{equation}
We refer the reader to the Appendix 1 and Appendix 2 for the explicit formulas defining 
amplitudes $|M|^2$ and $\Delta M$ in the non-relativistic approximations of the orders relevant in the analysis.

First we analyse how the normalized spectra depend on the the coupling phases.
From the results given in \ref{A1} we can read that for $\theta_{\rm S_R}=\pi/2$ 
(that is $\Re\left(c_{\rm S_R} d_{\rm S_R}\right)=0$) the amplitude $\Delta M_{\rm VA-{S_R}}$ 
reduces to $\Delta M_{\rm VA}$.
The leading terms of $|M_{\rm VA}|^2$ and $|M_{\rm VA-{S_R}}|^2$ are of the order O$(v^0,v_{\chi}^0)$ thus the 
plots of $(1/R) (dR/(dE_{\rm R}))$ on Fig. \ref{E_VASR_Dirac} do not discern the two cases,
whereas for $\theta_{\rm S_R}\neq\pi/2$ the differences are visible.

Second, we ask whether this scenario can help to determine the nature of DM.
It seems that the answer to this question is not affirmative: Fig. \ref{E_VASR_Majorana} displays 
only tiny differences between the Dirac and the Majorana DM in the corresponding cases.

Next, for tensor interactions with $\theta_{\rm T_R}=\pi/2$ the amplitude $\Delta M_{\rm VA-{T_R}}$ differs from $\Delta M_{\rm VA}$ by a nontrivial contribution from $\Delta M_{\rm T_R}$ which results in descending behaviour of $(1/|\Delta R|) (d|\Delta R|/(dE_{\rm R}))$ for small energies. This is shown on Fig. \ref{E_VATR_Dirac}. Similarly to the scalar case the leading term of $|M_{\rm VA-T_R}|^2$ is of the order O$(v^0,v_{\chi}^0)$ thus the normalized plot of $(1/R) (dR/(dE_{\rm R}))$ is the same as for $|M_{\rm VA}|^2$.

\begin{figure}[!h]
	\begin{center}
		\includegraphics[width=1\linewidth]{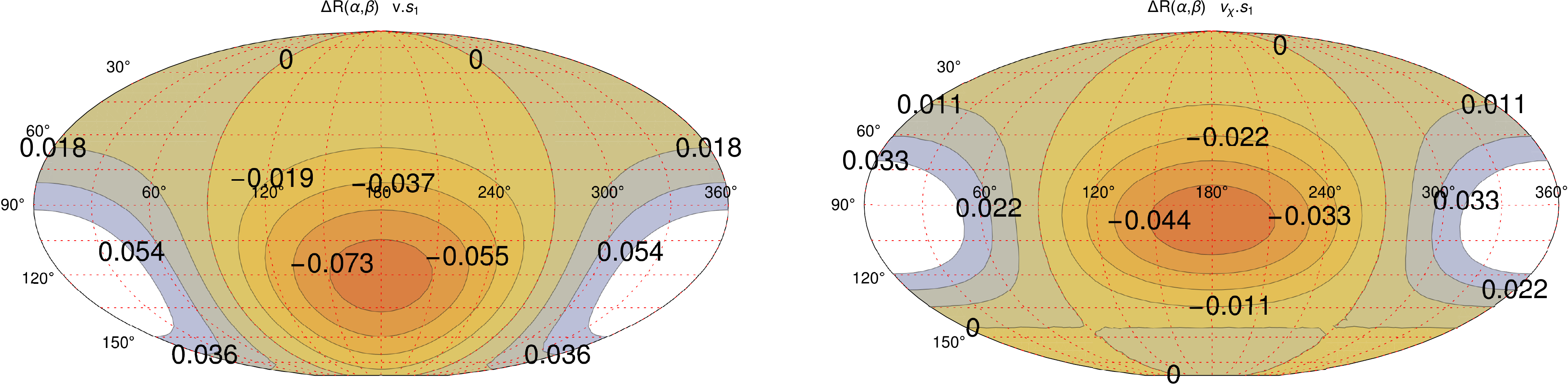}
	\end{center}
	\caption{Contributions of terms, $\vec{v}\cdot\hs{1}$, 
	$\vec{v}_{\chi}\cdot\hs{1}$, to  $(d^{2} R)/(d E_{\rm R}\,d\,\Omega)$ as a function of $ \alpha$ and $\beta$;
	incoming DM with $E_{\rm R}= 20$ keV.}
	\label{vs1vchis1}
\end{figure}

\begin{figure}[!h]
	\begin{center}
		\includegraphics[width=1\linewidth]{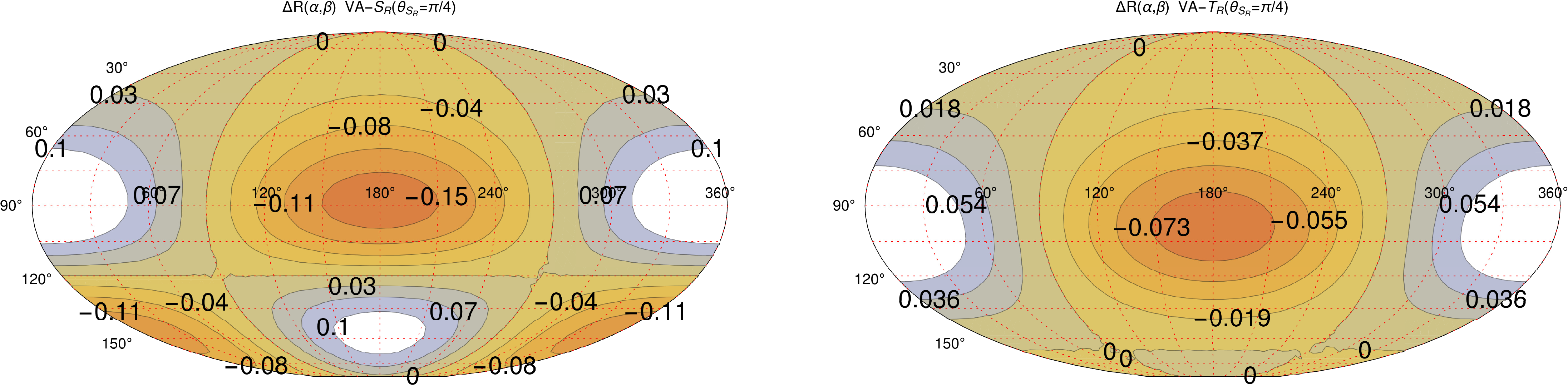}
	\end{center}
	\caption{Dirac DM for $ E_{\rm R}= 20$ keV:  plots of  $d^{2}\Delta R/(d E_{\rm R} d\,\Omega)$ as a function of $ \alpha$ and $\beta$; left plot for  V, A, S interactions with $\theta_{\rm S_{\rm R}}=\pi/4$;  right plot for  V, A, T interactions with  $\theta_{\rm S_{\rm R}}=\pi/4$. }
	\label{O_DVASRTR_thSpi4_thTpi4}
\end{figure}

\begin{figure}[!h]
	\begin{center}
		\includegraphics[width=1\linewidth]{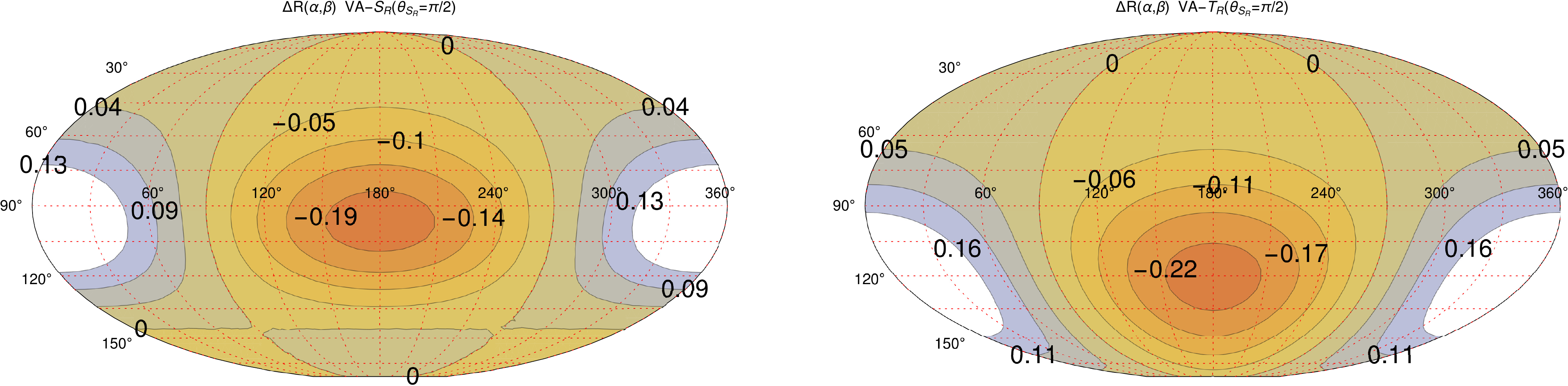}
	\end{center}
	\caption{Dirac DM for $ E_{\rm R}= 20$ keV:  plots of   $d^{2}\Delta R/(d E_{\rm R} d\,\Omega)$ as a function of $ \alpha$ and $\beta$; left plot for  V, A, S interactions with $\theta_{\rm S_{\rm R}}=\pi/2$;  right plot for  V, A, T interactions with  $\theta_{\rm S_{\rm R}}=\pi/2$.}
	\label{O_DVASRTR_thSpi2_thTpi2}
\end{figure}

Finally, we analyse the the angular distributions of recoil nucleus on $(\alpha, \beta)$-maps.
Fig. \ref{vs1vchis1} shows contributions of the two terms, $\vec{v}\cdot\hs{1}$, $\vec{v}_{\chi}\cdot\hs{1}$, 
to the detection rate $(d^{2} R)/(d E_{\rm R}\,d\,\Omega)$; the total detection rates in the discussed scenario are formed by superpositions of these dipol-like components.
Figs. \ref{O_DVASRTR_thSpi4_thTpi4} and  \ref{O_DVASRTR_thSpi2_thTpi2} present $(\alpha, \beta)$-charts 
of the detection rates $(d^{2} \Delta R)/(d E_{\rm R}\,d\,\Omega)$
for the scalar and tensor interactions. The plots reveal a small difference between the  ${\rm VA-S_R}$ and  
${\rm VA-T_R}$ scenarios. Moreover the shapes of the corresponding distributions are sensitive to the coupling phases. 

As an additional remark we state the $(\alpha, \beta)$-maps for the Majorana DM were analyzed as well; compared to the Dirac DM the difference is only in values (on the level $\sim60\%$), not in shapes.

It is worth pointing out that the possible effects of the time inversion symmetry are not visible because all triple mixed products made up of the momenta of the incoming DM and of the recoil nucleus, and the target polarization are strongly 
suppressed. The expected effects can only enter with the terms of the order O$(v^2,\;v^2_{\chi})$. 

\section{Second scenario - polarized target with polarized recoil nucleus}
\label{sec4}

This section is an extended version of the considerations in the previous one. In addition, it is assumed that the polarization of the recoil nucleus will be measured. As a result,  various  mixed  products among the target polarization vector, the polarization vector of the recoil nucleus, the momenta of  incoming DM and of  recoil nucleus appear in the cross section. 
This indicates a breaking symmetry with respect to time inversion. 

\begin{figure}[!h]
	\begin{center}
		\includegraphics[width=0.68\linewidth]{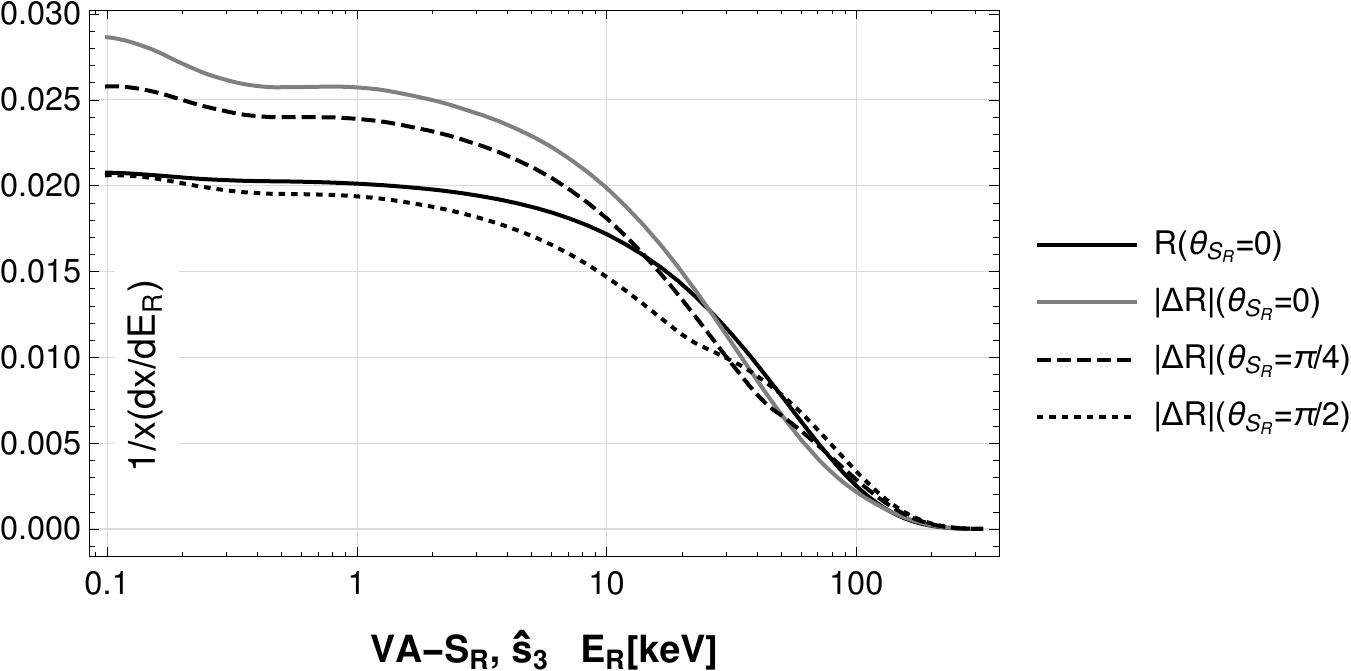}
	\end{center}
	\caption{Dirac DM with  polarization of recoil nucleus $\hs{3}$ for $\varphi_3 = \pi/2$, $\theta_3 = \pi/2$: plot of  $(1/R) (dR/d E_{\rm R})$ as a function of $ E_{\rm R}$ for vector V, axial A and scalar S interactions with $\theta_{\rm S_{\rm R}}=0$  (solid black line); plots of $(1/|\Delta R|) (d|\Delta R|/d E_{\rm R})$ as a function of $ E_{\rm R}$   for V, A, S interactions; solid gray line for $\theta_{\rm S_{\rm R}}=0$; dashed line for $\theta_{\rm S_{\rm R}}=\pi/4$; dotted line for $\theta_{\rm S_{\rm R}}=\pi/2$. }
	\label{E_VASRs3}
\end{figure}

\begin{figure}[!h]
	\begin{center}
		\includegraphics[width=1\linewidth]{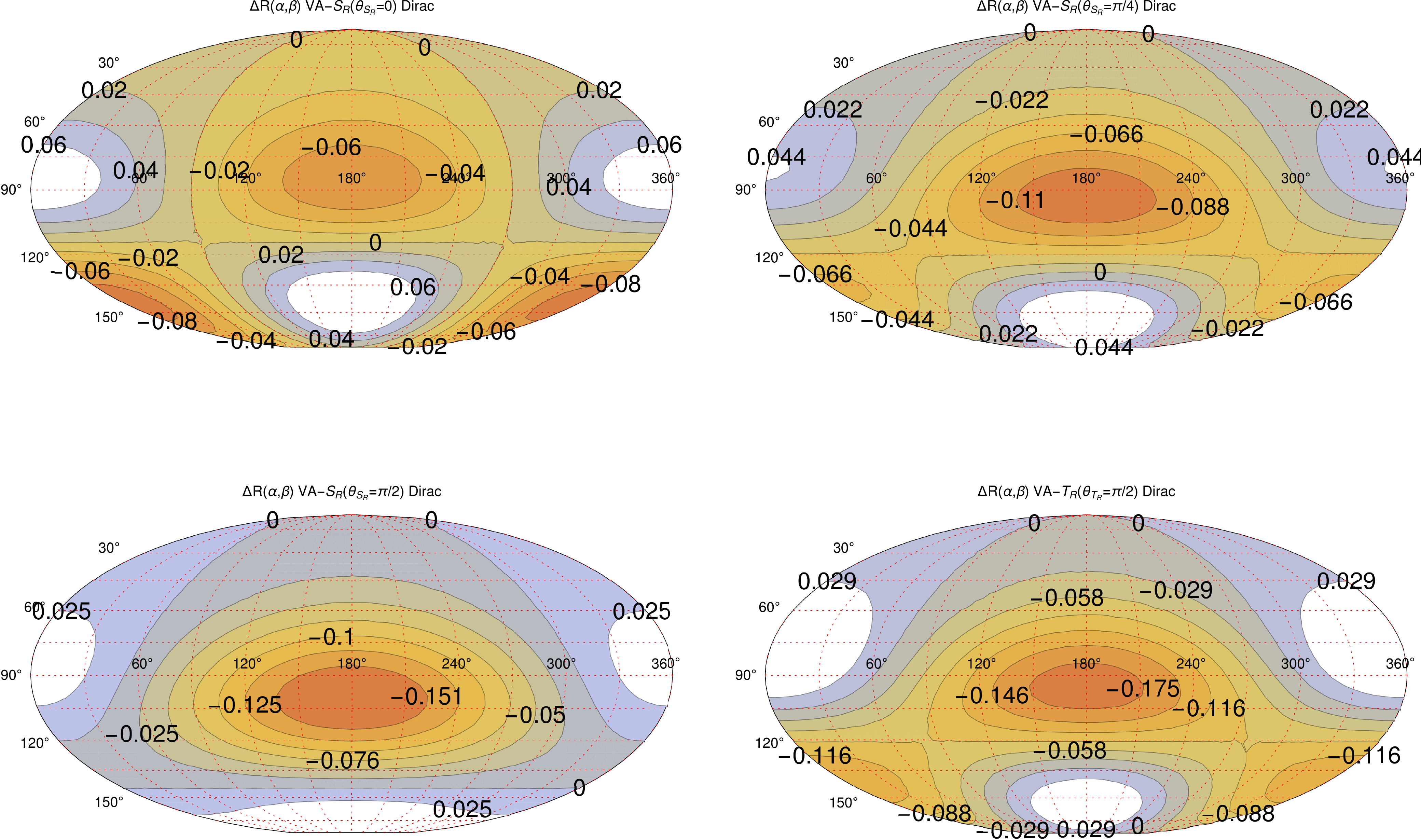}
	\end{center}
	\caption{Dirac DM with  polarization of recoil nucleus $\hs{3}$ for $\varphi_3 = \pi/2$, $\theta_3 = \pi/2$, $ E_{\rm R}= 20$ keV:  plots of  $d^{2}\Delta R/(d E_{\rm R} d\,\Omega)$ as a function of $ \alpha$ and $\beta$; top left plot for  V, A, S interactions with $\theta_{\rm S_{\rm R}}=0$; top right plot for  V, A, S interactions with 
		$\theta_{\rm S_{\rm R}}=\pi/4$;  bottom left plot for  V, A, S interactions with  $\theta_{\rm S_{\rm R}}=\pi/2$; bottom right plot for  V, A, T interactions with  $\theta_{\rm S_{\rm R}}=\pi/2$.}
		\label{O_DVASTRs3}
	\end{figure}

From the formulas for $|M_{\rm VA}|^2$ and $\Delta M_{\rm VA}$ we see that in the leading order there is no dependence on the $\hs{3}$, thus the spectral and angular distributions are insensitive to the polarization recoil nucleus. 
With the presence of the scalar interaction the leading order terms in $|M_{\rm S_R}(\hs{1},\hs{3})|^2$, $\Re(M_{\rm VA-S_R}(\hs{1},\hs{3}))$ and $\Delta M_{\rm S_R}(\hs{1},\hs{3})$, $\Re(\Delta M_{\rm VA-S_R}(\hs{1},\hs{3}))$ have the form, $\hs{1}\cdot\hs{3}$.
Therefore unless the recoil nucleus polarization is orthogonal to the target polarization the normalized spectra
of $R$ and $\Delta R$ will coincide for each $\hs{3}$ orientation.
Thus to observe a potential $\hs{3}$-dependent effects in what follows we specify the $\hs{1}\cdot\hs{3}=0$ orientation. 

As in the first scenario, the normalized spectra are sensitive to the coupling phases. 
But now the dependence is modified due to the presence of the new terms including the polarization of the outgoing nucleus. Fig. \ref{E_VASRs3} and \ref{O_DVASTRs3} were plotted for a specific orientation, $\hs{3} = \left(\cos{\varphi_3}\sin{\theta_3}, \sin{\varphi_3}\sin{\theta_3},\cos{\theta_3}\right)$
with $\varphi_3 = \pi/2$, $\theta_3 = \pi/2$ (that is $\hs{3}=\hat{y}$). The plots manifest a phase-dependent behaviour and in addition point to the different patterns for the scalar and tensor interactions. 
Interestingly one can observe CP breaking effects directly on the angular distribution plots in 
Fig. \ref{O_DVASTRs3}. To see them let us first notice that the kinematics of the process implies that the 
terms $\vec{v}_{\chi}\cdot\hs{1}$ and $\vec{v}\cdot\hs{1}$ in the amplitudes do not contribute to the 
differential event rates $(d^2 R)/(d\,E_{\rm R}\,d\Omega)$ for the nuclei recoiled in the direction 
orthogonal to $\hs{1}$. As a result one can see a ``null" circle along the lines of constant 
$\beta$ equal to $\pi/2$ and $3/2\,\pi$ in the Fig. \ref{vs1vchis1} and on all $(\alpha, \beta)$-charts of 
the first scenario. 
In the present case this feature is visible on the map with the phase $\theta_{\rm S_R}=0$ in Fig. \ref{O_DVASTRs3},
and is disturbed by the CP breaking effects induced by the mixed products as is seen on the remaining three maps in Fig. \ref{O_DVASTRs3} for the non-zero coupling phases \footnote{Notice that this characteristic can also be destroyed by the $\hs{1}\cdot\hs{3}$ term but in the discussed case it is assumed to be zero.}.

Furthermore, we can ask whether it is possible to find explicit observables isolating 
CP breaking terms. In fact, the assumed orientation of $\hs{1}$ and $\hs{3}$ allows us to define quantities
\begin{eqnarray}
 {\rm D_{(1)}} M_{\rm VA-S_R}(\hs{1},\hs{3}) &:=& \Delta M_{\rm VA-S_R}(\hs{1},\hs{3}) - \Delta M_{\rm VA-S_R}(\hs{1},-\hs{3}) \\
 {\rm D_{(1)}} M_{\rm VA-T_R}(\hs{1},\hs{3}) &:=& \Delta M_{\rm VA-T_R}(\hs{1},\hs{3}) - \Delta M_{\rm VA-T_R}(\hs{1},-\hs{3})
\end{eqnarray}
containing exclusively $\hat{v}_{\chi }\cdot (\hs{1}\times \hs{3})$ and $\hat{v}\cdot (\hs{1}\times \hs{3})$
terms. On the angular distribution maps each of the mixed products would give contours lying along lines of constant $\alpha$. 
Another choice is
\begin{eqnarray}
 {\rm D_{(2)}} M_{\rm VA-S_R}(\hs{1},\hs{3},\vec{v}) &:=& \Delta M_{\rm VA-S_R}(\hs{1},\hs{3},\vec{v}) - \Delta M_{\rm VA-S_R}(\hs{1},\hs{3},-\vec{v}), \label{eq_s1s3v_ST}\\
 {\rm D_{(2)}} M_{\rm VA-T_R}(\hs{1},\hs{3},\vec{v}) &:=& \Delta M_{\rm VA-T_R}(\hs{1},\hs{3},\vec{v}) - \Delta M_{\rm VA-T_R}(\hs{1},\hs{3},-\vec{v})\nonumber
\end{eqnarray}
for $\hat{v}\cdot\hs{1}=0$ leaving only the term $\hat{v}\cdot (\hs{1}\times \hs{3})$.

\section{Third scenario - polarized target  with polarized incoming dark matter}
\label{sec5}

In this section, the physical consequences of the assumption that DM is polarized are discussed. In this case the polarization of the recoil nucleus is not measured, but the symmetry violation under the time inversion generated by the  triple mixed products among the target polarization vector, the polarization vector of incoming DM, the momenta of  incoming DM and of  recoil nucleus may occur. 

From the formulas $|M(\hs{1},\hs{2})|^2$ and $\Delta M(\hs{1},\hs{2})$ for the vector, scalar and tensor interactions we see that the dominating contribution comes from the scalar product $\hs{1}\cdot\hs{2}$. This term on the angular distribution maps would produce contours along lines of constant $\alpha$ (reflecting behaviour of the integral kernel used in \ref{r3}); the overall scale would depend on a mutual orientation between $\hs{1}$ and $\hs{2}$. 
At this point it is worth to note that the presence of DM polarization makes the detection rates larger by the factor of the order of $\sim10^3$ as compared to the scenario without polarization. This also is true in the absence of scalar and tensor interactions.
\begin{figure}[!h]
	\begin{center}
		\includegraphics[width=1\linewidth]{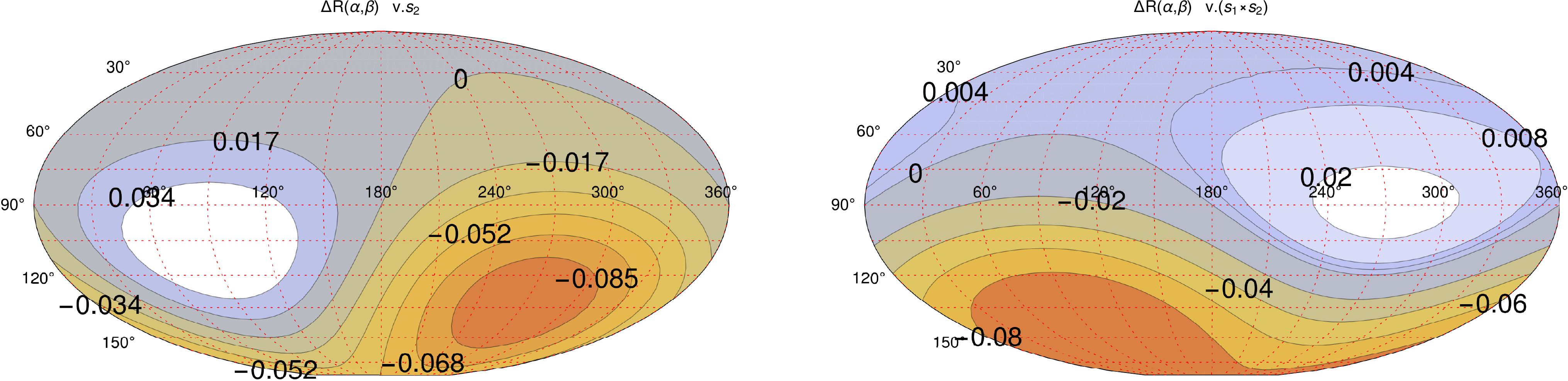}
	\end{center}
	\caption{
	Contributions of terms, $\vec{v}\cdot\hs{2}$, $\vec{v}\cdot(\hs{1}\times\hs{2})$, to  
	$(d^{2} R)/(d E_{\rm R}\,d\,\Omega)$ as a function of $ \alpha$ and $\beta$;
	incoming DM with $\kappa = \pi/3$, $\zeta=\pi/2$, $E_{\rm R}= 20$ keV.}
	\label{vs2vs1s2}
\end{figure}
Theoretically one can even choose an orientation of the polarized target to eliminate this predominant term
to observe higher order effects. Fig. \ref{vs2vs1s2} shows the two contributions, one from 
P-breaking term $\vec{v}\cdot\hs{2}$, and second, from P and CP-breaking term $\vec{v}\cdot(\hs{1}\times\hs{2})$.
The DM polarization is there parameterized as $\hs{2} = \left(\cos{\zeta}\sin{\kappa}, \sin{\zeta}\sin{\kappa},\cos{\kappa}\right)$; on both maps $\kappa=\pi/3$, $\zeta=\pi/2$ and $\hs{1}=(1,0,0)$. Similarly to Fig. \ref{vs1vchis1} the present maps have dipolar characteristics, but now with respect to different axes whose orientation depends on $\hs{2}$. The analogous (not presented) contributions coming from the terms, $\vec{v}_{\chi}\cdot\hs{2}$ and $\vec{v}_{\chi}\cdot(\hs{1}\times\hs{2})$, have comparable shapes.

\section{Polarized target  with polarized incoming dark matter and polarized recoil nucleus}
\label{sec6}

The most general and at the same time quite exotic case assumes that apart from the polarized target, the incoming DM can be polarized and the polarization of the recoil nucleus  is measured. This means that the effects of time inversion symmetry violation   can be caused by triple mixed products built only from the polarization vectors themselves. 

In the present scenario the leading terms of $|M(\hs{1},\hs{2},\hs{3})|^2$ are 
of the order O$(v_{\chi}^0,v^0)$; they read:
\begin{eqnarray}
|M_{\rm VA-S_R}(\hs{1},\hs{2},\hs{3})|^2 &=& 
2 m^2 \left(4 |c_{\rm S_R}| {}^2 |d_{\rm S_R}| {}^2 + 2 \hat{s}_1\cdot (\hat{s}_2\times \hat{s}_3)
\Im\left(c_{\rm S_R} d_{\rm S_R}\right) - 2 \hat{s}_2\cdot \hat{s}_3 \Re\left(c_{\rm S_R} d_{\rm S_R}\right)  \right. \nonumber\\
&&\left. +2 \Re\left(c_{\rm S_R} d_{\rm S_R}\right)
-\hat{s}_1\cdot \hat{s}_2 \left(2 \Re\left(c_{\rm S_R} d_{\rm S_R}\right)+1\right)+1\right) m_{\chi }^2 \\
|M_{\rm VA-T_R}(\hs{1},\hs{2},\hs{3})|^2 &=& 
2 m^2 \left(4 | c_{\rm T_R}| {}^2 \left(-2 \hat{s}_1\cdot \hat{s}_2+2 \hat{s}_2\cdot \hat{s}_3+3\right) 
| d_{\rm T_R}| {}^2+2 \hat{s}_1\cdot \hat{s}_2\times \hat{s}_3 \Im\left(c_{\rm T_R} d_{\rm T_R}\right)   \right. \nonumber\\
&&\left. -2 \hat{s}_2\cdot 
\hat{s}_3 \Re\left(c_{\rm T_R} d_{\rm T_R}\right)-6 \Re\left(c_{\rm T_R} d_{\rm T_R}\right)+\hat{s}_1\cdot \hat{s}_2 \left(6 \Re\left(c_{\rm T_R} d_{\rm T_R}\right)-1\right)+1\right) m_{\chi }^2
\end{eqnarray}
The above formulae assume that the polarization of the recoil nucleus is perpendicular to the polarization of the target. We can construct two new observables to rule out all the double products   
$\Delta_{\rm S}:= |M_{\rm VA-S_R}|^2(\hs{1},\hs{2},\hs{3})+|M_{\rm VA-S_R}|^2(-\hs{1},\hs{2},-\hs{3})$, 
and $\Delta_{\rm T}$ in the analogous way; in the assumed order they read:
\begin{eqnarray}
\Delta_{\rm S} & = & 4 m^2 \left(4 | c_{\rm S_R}| {}^2 | d_{\rm S_R}| {}^2+2 \hat{s}_1\cdot (\hat{s}_2\times \hat{s}_3) \Im\left(c_{\rm S_R} d_{\rm S_R}\right)+2 \Re\left(c_{\rm S_R} d_{\rm S_R}\right)+1\right) m_{\chi }^2 \label{eq_s1s2s3_ST}\\
\Delta_{\rm T} & = & 4 m^2 \left(12 | c_{\rm T_R}| {}^2 | d_{\rm T_R}| {}^2+2 \hat{s}_1\cdot (\hat{s}_2\times \hat{s}_3) \Im\left(c_{\rm T_R} d_{\rm T_R}\right)-6 \Re\left(c_{\rm T_R} d_{\rm T_R}\right)+1\right) m_{\chi }^2\nonumber
\end{eqnarray}

We see that the detection rate depends now on the orientation between $\hs{1}\times\hs{3}$ and DM polarization 
$\hs{2}$. Any change of the number of events that appears during rigid rotations of the pair $\hs{1}$, $\hs{3}$ would signal time reversal violation (i.e. presence of $\hat{s}_1\cdot (\hat{s}_2\times \hat{s}_3) \Im\left(c_{\rm S_R} d_{\rm S_R}\right)\sim \hs{1}\cdot(\hs{2}\times\hs{3}) \,\sin{(\theta_{\rm S_{\rm R}}})$ term). 
 %\item
Moreover, rigid rotations of $\hs{1}$, $\hs{3}$ in principle would allow for determination of the DM polarization direction. 
From theoretical point of view the scenario analyzed is rather peculiar because it primarily assumes a mere presence 
of DM polarization. In addition, an experiment would require a control over polarization of the recoil nucleus. 

\section{Conclusions}
\label{sec7} 

We have analyzed the theoretically possible scenarios for the fermionic dark matter interacting 
with the target of polarized nuclei in the presence of non-standard vector, axial, scalar and tensor 
interactions. We have specified to the two kinds of measurements: normalized spectra and angular distribution of the recoil nucleus. We have considered four experiments with polarized target. In the first we assumed  
that the incoming DM is unpolarized and the spin of the recoiled nucleus is not measured. In the second 
option we allow for the recoiled spin measurement. The third scenario permits a hypothetical polarization
of DM; in this case the recoiled nucleus polarization is not registered. In the fourth experiment 
we assume both, DM polarization and the recoiled nucleus spin detection.

The analysis presented in the paper shows that each scenario offers new ways to study a nature of DM
and its interaction with SM particles. It is reflected in the dependence of the detection rates on the coupling constants and orientations of the spin polarizations. This feature is seen on both, spectral diagrams and 
angular distribution maps. The parameter space of the considered models is rather large and to make general conclusions a complete analysis would be required which goes beyond the cases considered. However even in the subset 
of presented models we can observe some regularities in the normalized spectra as is seen in Figs. 
(\ref{E_VASR_Dirac}, \ref{E_VASR_Majorana}, \ref{E_VATR_Dirac}, \ref{E_VASRs3}). On the other hand we can also notice an unexpected behaviour of the spectrum at low energies for tensor interactions 
(see Fig. \ref{E_VATR_Dirac}). Next, we would like to point out an important role that the angular distributions 
may play in studies of DM nature. The analysed examples have shown that $(\alpha, \beta)$-maps could be useful
in testing time reversal symmetry breaking effects, see Figs. (\ref{O_DVASTRs3}, \ref{vs2vs1s2}). In particular one can define observables which isolate CP-breaking interaction terms, Eqs. (\ref{eq_s1s3v_ST}, \ref{eq_s1s2s3_ST}). 
In the more exotic scenarios with polarized DM 
one can, in principle, put forward a measurement scheme determining DM spin orientation, see Sec. {\ref{sec5}}. 

The great problem is the lack of any experiments with a polarized target, so this and previous works of other authors have been carried out in the hope that they will encourage research groups to take action in this direction. 

\section*{ Declaration of competing interest}

The authors declare that they have no known competing financial interests or personal relationships that could have appeared
to influence the work reported in this paper. 
\section*{Acknowledgment}
This research did not receive any specific grant from funding agencies in the public, commercial, or
not-for-profit sectors.

\section*{Appendix 1: Squared modulus of amplitudes for DM scattering  on polarized target}
\label{A1} 

\noindent
$\hs{1}\neq0,\;\hs{2}=0,\;\hs{3}=0$ case, terms to the order O$(v,v_{\chi})$:
\begin{eqnarray*}
	|M_{\rm VA}(\hs{1})|^2 &=& 4 m^2 m_{\chi }^2 \left(\hat{v}_{\chi }\cdot \hat{s}_1 v_{\chi }+1\right) \\
	|M_{\rm S_R}(\hs{1})|^2 &=& 16 m^2 \left| c_{\rm S_R}\right| {}^2 \left| d_{\rm S_R}\right| {}^2 m_{\chi }^2 \\
	|M_{\rm T_R}(\hs{1})|^2 &=& -16 m \left| c_{\rm T_R}\right| {}^2 \left| d_{\rm T_R}\right| {}^2 m_{\chi } \left(m \left(m_{\chi } 
	\left(4 \hat{v}_{\chi }\cdot \hat{s}_1 v_{\chi }-3\right)-2 m v \hat{v}\cdot \hat{s}_1\right)-
	m v \hat{v}\cdot \hat{s}_1 m_{\chi }\right) \\
	2\Re[M_{\rm VA-S_R}(\hs{1})] &=& 4 m \Re\left(c_{\rm S_R} d_{\rm S_R}\right) m_{\chi } \left(2 m \left(-m v \hat{v}\cdot 
	\hat{s}_1+m_{\chi }+\hat{v}_{\chi }\cdot \hat{s}_1 m_{\chi } v_{\chi }\right)-m v \hat{v}\cdot \hat{s}_1 m_{\chi }\right) \\
	2\Re(M_{\rm VA-T_R}(\hs{1})) &=& 8 m \Re\left(c_{\rm T_R} d_{\rm T_R}\right) m_{\chi } \left(m \left(m_{\chi } \left(\hat{v}_{\chi }\cdot \hat{s}_1 v_{\chi }-3\right)-m v \hat{v}\cdot \hat{s}_1\right)-m v \hat{v}\cdot \hat{s}_1 m_{\chi }\right) \\
	|M_{\rm VA}^{\rm Maj}(\hs{1})|^2 &=& 4 m \, m_{\chi } \left(m v \hat{v}\cdot \hat{s}_1 m_{\chi }+m \left(m_{\chi } \left(2 \hat{v}_{\chi }\cdot \hat{s}_1 v_{\chi }+3\right)-m v \hat{v}\cdot \hat{s}_1\right)\right) \\
	|M_{\rm S_R}^{\rm Maj}(\hs{1})|^2 &=& 64 m^2 \left| c_{\rm S_R}\right| {}^2 \left| d_{\rm S_R}\right| {}^2 m_{\chi }^2 \\
	2\Re(M_{\rm VA-S_R}^{\rm Maj}(\hs{1})) &=& -16 m^3 v \hat{v}\cdot \hat{s}_1 \Re\left(c_{\rm S_R} d_{\rm S_R}\right) m_{\chi }
	\end{eqnarray*}
	\noindent
	$\hs{1}\neq0,\;\hs{2}=0,\;\hs{3}\neq0$ case, terms to the order O$(v,v_{\chi})$:
	\begin{eqnarray*}
	|M_{\rm VA}(\hs{1},\hs{3})|^2 &=& 2 m \, m_{\chi } \left(m \left(-m v \hat{v}\cdot \hat{s}_3+m_{\chi }+\hat{v}_{\chi }\cdot \hat{s}_1 m_{\chi } v_{\chi }+\hat{v}_{\chi }\cdot \hat{s}_3 m_{\chi } v_{\chi }\right)-m v \hat{v}\cdot \hat{s}_3 m_{\chi }\right) \\
	|M_{\rm S_R}(\hs{1},\hs{3})|^2  &=& 8 m^2 \left| c_{\rm S_R}\right| {}^2 \left| d_{\rm S_R}\right| {}^2 \left(\hat{s}_1\cdot \hat{s}_3+1\right)  m_{\chi }^2 \\
	|M_{\rm T_R}(\hs{1},\hs{3})|^2  &=& -8 m \left| c_{\rm T_R}\right| {}^2 \left| d_{\rm T_R}\right| {}^2 m_{\chi } \left(m v \left(\hat{v}\cdot \hat{s}_1-3 \hat{v}\cdot \hat{s}_3\right) m_{\chi }+m \left(2 m v \left(\hat{v}\cdot \hat{s}_1-\hat{v}\cdot \hat{s}_3\right)\right.\right. \\
	&&\quad \left.\left. + m_{\chi } \left(\hat{s}_1\cdot \hat{s}_3-4 \hat{v}_{\chi }\cdot \hat{s}_1 v_{\chi }+4 \hat{v}_{\chi }\cdot \hat{s}_3 v_{\chi }-3\right)\right)\right) \\
	2\Re(M_{\rm VA-S_R}(\hs{1},\hs{3})) &=& 2 m \, m_{\chi } \left(\Im\left(c_{\rm S_R} d_{\rm S_R}\right) \left(2 m \hat{v}_{\chi }\cdot (\hat{s}_1\times \hat{s}_3) m_{\chi } v_{\chi }-m v \hat{v}\cdot (\hat{s}_1\times \hat{s}_3) \left(2 m+m_{\chi }\right)\right)\right. \\
	&&\quad \left. -\Re\left(c_{\rm S_R} d_{\rm S_R}\right) \left(m v \left(\hat{v}\cdot \hat{s}_1+\hat{v}\cdot \hat{s}_3\right) m_{\chi }-2 m \left(m_{\chi } \left(\hat{s}_1\cdot \hat{s}_3+\hat{v}_{\chi }\cdot \hat{s}_1 v_{\chi }+\hat{v}_{\chi }\cdot \hat{s}_3 v_{\chi }+1\right) \right.\right.\right. \\
	&&\quad \left.\left.\left. - m v \left(\hat{v}\cdot \hat{s}_1+\hat{v}\cdot \hat{s}_3\right)\right)\right)\right) \\
	2\Re(M_{\rm VA-T_R}(\hs{1},\hs{3})) &=& 4 m \, m_{\chi } \left(m \Im\left(c_{\rm T_R} d_{\rm T_R}\right) \left(\hat{v}_{\chi }\cdot (\hat{s}_1\times \hat{s}_3) m_{\chi } v_{\chi }-m v \hat{v}\cdot (\hat{s}_1\times \hat{s}_3)\right)\right. \\
	&&\quad \left. -\Re\left(c_{\rm T_R} d_{\rm T_R}\right) \left(m v \left(\hat{v}\cdot \hat{s}_1-3 \hat{v}\cdot \hat{s}_3\right) m_{\chi }+m \left(m v \left(\hat{v}\cdot \hat{s}_1-3 \hat{v}\cdot \hat{s}_3\right) \right.\right.\right. \\
	&&\quad \left.\left.\left. - m_{\chi } \left(\hat{s}_1\cdot \hat{s}_3+\hat{v}_{\chi }\cdot \hat{s}_1 v_{\chi }-3 \hat{v}_{\chi }\cdot \hat{s}_3 v_{\chi }-3\right)\right)\right)\right)
	\end{eqnarray*}
	\noindent
$\hs{1}\neq0,\;\hs{2}\neq0,\;\hs{3}=0$ case, terms to the order O$(v,v_{\chi})$:
	\begin{eqnarray*}
	|M_{\rm VA}(\hs{1},\hs{2})|^2 &=& -4 m^2 m_{\chi }^2 \left(\hat{s}_1\cdot \hat{s}_2-\hat{v}_{\chi }\cdot \hat{s}_1 v_{\chi }+
	\hat{v}_{\chi }\cdot \hat{s}_2 v_{\chi }-1\right) \\
	|M_{\rm S_R}(\hs{1},\hs{2})|^2 &=& 16 m^2 \left| c_{\rm S_R}\right| {}^2 \left| d_{\rm S_R}\right| {}^2 m_{\chi } \left(m v \hat{v}\cdot \hat{s}_2+m_{\chi }\right) \\
	|M_{\rm T_R}(\hs{1},\hs{2})|^2 &=& 16 m \left| c_{\rm T_R}\right| {}^2 \left| d_{\rm T_R}\right| {}^2 m_{\chi } \left(m v \left(\hat{v}
	\cdot \hat{s}_1-2 \hat{v}\cdot \hat{s}_2\right) m_{\chi } \right. \\
	&&\quad \left. + m \left(m_{\chi } \left(-2 \hat{s}_1\cdot \hat{s}_2-4 \hat{v}_{\chi }\cdot \hat{s}_1 v_{\chi }+4 \hat{v}_{\chi }\cdot \hat{s}_2 v_{\chi }+3\right)-m v \left(\hat{v}\cdot \hat{s}_2-2 \hat{v}\cdot \hat{s}_1\right)\right)\right) \\
	2\Re(M_{\rm VA-S_R}(\hs{1},\hs{2})) &=& 4 m \, m_{\chi } \left(-m v \hat{v}\cdot (\hat{s}_1\times \hat{s}_2) \Im\left(c_{\rm S_R} d_{\rm S_R}\right) \left(2 m+m_{\chi }\right)-\Re\left(c_{\rm S_R} d_{\rm S_R}\right) \left(m v \left(\hat{v}\cdot \hat{s}_1-\hat{v}\cdot \hat{s}_2\right) m_{\chi } \right.\right. \\
	&&\quad \left.\left. + 2 m \left(m v \left(\hat{v}\cdot \hat{s}_1-\hat{v}\cdot \hat{s}_2\right)+m_{\chi } \left(\hat{s}_1\cdot \hat{s}_2-\hat{v}_{\chi }\cdot \hat{s}_1 v_{\chi }+\hat{v}_{\chi }\cdot \hat{s}_2 v_{\chi }-1\right)\right)\right)\right) \\
	2\Re(M_{\rm VA-T_R}(\hs{1},\hs{2})) &=& -8 m \, m_{\chi } \left(\Im\left(c_{\rm T_R} d_{\rm T_R}\right) \left(m v \hat{v}\cdot (\hat{s}_1\times \hat{s}_2) \left(m+m_{\chi }\right)-4 m \hat{v}_{\chi }\cdot (\hat{s}_1\times \hat{s}_2) m_{\chi } v_{\chi }\right)\right.\\
	&&\quad \left. - \Re\left(c_{\rm T_R} d_{\rm T_R}\right) \left(m v \left(\hat{v}\cdot \hat{s}_2-\hat{v}\cdot \hat{s}_1\right) m_{\chi }+m \left(m v \left(\hat{v}\cdot \hat{s}_2-\hat{v}\cdot \hat{s}_1\right)\right.\right.\right. \\
	&&\quad\left.\left.\left. + m_{\chi } \left(3 \hat{s}_1\cdot \hat{s}_2+\hat{v}_{\chi }\cdot \hat{s}_1 v_{\chi }-\hat{v}_{\chi }\cdot \hat{s}_2 v_{\chi }-3\right)\right)\right)\right)
	\end{eqnarray*}
	\noindent
	$\hs{1}\neq0,\;\hs{2}\neq0,\;\hs{3}\neq0$ case, terms to the order O$(v^0,v_{\chi}^0)$:
	\begin{eqnarray*}
	|M_{\rm VA}(\hs{1},\hs{2},\hs{3})|^2 &=& -2 m^2 \left(\hat{s}_1\cdot \hat{s}_2-1\right) m_{\chi }^2 \\
	|M_{\rm S_R}(\hs{1},\hs{2},\hs{3})|^2 &=& 8 m^2 \left| c_{\rm S_R}\right| {}^2 \left| d_{\rm S_R}\right| {}^2 \left(\hat{s}_1\cdot \hat{s}_3+1\right) m_{\chi }^2 \\
	|M_{\rm T_R}(\hs{1},\hs{2},\hs{3})|^2 &=& 8 m^2 \left| c_{\rm T_R}\right| {}^2 \left| d_{\rm T_R}\right| {}^2 \left(-2 \hat{s}_1\cdot \hat{s}_2-\hat{s}_1\cdot \hat{s}_3+2 \hat{s}_2\cdot \hat{s}_3+3\right) m_{\chi }^2 \\
	2\Re(M_{\rm VA-S_R}(\hs{1},\hs{2},\hs{3})) &=& 4 m^2 \left(\hat{s}_1\cdot (\hat{s}_2\times \hat{s}_3) \Im\left(c_{\rm S_R} d_{\rm S_R}\right)-\left(\hat{s}_1\cdot \hat{s}_2-\hat{s}_1\cdot \hat{s}_3+\hat{s}_2\cdot \hat{s}_3-1\right) 
	\Re\left(c_{\rm S_R} d_{\rm S_R}\right)\right) m_{\chi }^2 \\
	2\Re(M_{\rm VA-T_R}(\hs{1},\hs{2},\hs{3})) &=& 4 m^2 \left(\hat{s}_1\cdot (\hat{s}_2\times \hat{s}_3) \Im\left(c_{\rm T_R} d_{\rm T_R}\right)-\left(-3 \hat{s}_1\cdot \hat{s}_2-\hat{s}_1\cdot \hat{s}_3+\hat{s}_2\cdot \hat{s}_3+3\right) \Re\left(c_{\rm T_R} d_{\rm T_R}\right)\right) m_{\chi }^2 
\end{eqnarray*}

\section*{Appendix 2:  Difference of squared modulus of amplitudes for DM scattering on polarized target}
\label{A2}

\noindent
$\hs{1}\neq0,\;\hs{2}=0,\;\hs{3}=0$ case, terms to the order O$(v,v_{\chi})$:
\begin{eqnarray*}
	\Delta M_{\rm VA}(\hs{1}) &=& 4 m^2 \hat{v}_{\chi }\cdot \hat{s}_1 m_{\chi }^2 v_{\chi } \\
	\Delta M_{\rm S_R}(\hs{1}) &=& 0 \\
	\Delta M_{\rm T_R}(\hs{1}) &=& 16 m \left| c_{\rm T_R}\right| {}^2 \left| d_{\rm T_R}\right| {}^2 m_{\chi } \left(2 v \hat{v}\cdot \hat{s}_1 m^2+v \hat{v}\cdot \hat{s}_1 m_{\chi } m-4 \hat{v}_{\chi }\cdot \hat{s}_1 m_{\chi } v_{\chi } m\right) \\
	2\Re(\Delta M_{\rm VA-S_R}(\hs{1})) &=& -4 m \Re\left(c_{\rm S_R} d_{\rm S_R}\right) m_{\chi } \left(2 v \hat{v}\cdot \hat{s}_1 m^2+v \hat{v}\cdot \hat{s}_1 m_{\chi } m-2 \hat{v}_{\chi }\cdot \hat{s}_1 m_{\chi } v_{\chi } m\right) \\
	2\Re(\Delta M_{\rm VA-T_R}(\hs{1})) &=& 8 m \Re\left(c_{\rm T_R} d_{\rm T_R}\right) m_{\chi } \left(-v \hat{v}\cdot \hat{s}_1 m^2-v \hat{v}\cdot \hat{s}_1 m_{\chi } m+\hat{v}_{\chi }\cdot \hat{s}_1 m_{\chi } v_{\chi } m\right) \\
	\Delta M_{\rm VA}^{\rm Maj}(\hs{1}) &=& 4 m \, m_{\chi } \left(-v \hat{v}\cdot \hat{s}_1 m^2+v \hat{v}\cdot \hat{s}_1 m_{\chi } m+2 \hat{v}_{\chi }\cdot \hat{s}_1 m_{\chi } v_{\chi } m\right) \\
	\Delta M_{\rm VA}^{\rm Maj}(\hs{1}) &=& 0 \\
	2\Re(\Delta M_{\rm VA-S_R}^{\rm Maj}(\hs{1})) &=& -16 m^3 v \hat{v}\cdot \hat{s}_1 \Re\left(c_{\rm S_R} d_{\rm S_R}\right) m_{\chi }
	\end{eqnarray*}
	\noindent
	$\hs{1}\neq0,\;\hs{2}=0,\;\hs{3}\neq0$ case, terms to the order O$(v,v_{\chi})$:
	\begin{eqnarray*}
	\Delta M_{\rm VA}(\hs{1},\hs{3}) &=& 2 m^2 \hat{v}_{\chi }\cdot \hat{s}_1 m_{\chi }^2 v_{\chi } \\
	\Delta M_{\rm SR}(\hs{1},\hs{3}) &=& 8 m^2 \left| c_{\rm S_R}\right| {}^2 \left| d_{\rm S_R}\right| {}^2 \hat{s}_1\cdot \hat{s}_3 m_{\chi }^2 \\
	\Delta M_{\rm T_R}(\hs{1},\hs{3}) &=& -8 m \left| c_{\rm T_R}\right| {}^2 \left| d_{\rm T_R}\right| {}^2 m_{\chi } \left(m v \hat{v}\cdot \hat{s}_1 m_{\chi }+m \left(2 m v \hat{v}\cdot \hat{s}_1+m_{\chi } \left(\hat{s}_1\cdot \hat{s}_3-4 \hat{v}_{\chi }\cdot \hat{s}_1 v_{\chi }\right)\right)\right) \\
	2\Re(\Delta M_{\rm VA-S_R}(\hs{1},\hs{3})) &=& -2 m \, m_{\chi } \left(\Im\left(c_{\rm S_R} d_{\rm S_R}\right) \left(m v \hat{v}\cdot (\hat{s}_1\times \hat{s}_3) \left(2 m+m_{\chi }\right)-2 m \hat{v}_{\chi }\cdot (\hat{s}_1\times \hat{s}_3) m_{\chi } v_{\chi }\right) \right. \\
	&&\quad \left. + \Re\left(c_{\rm S_R} d_{\rm S_R}\right) \left(m v \hat{v}\cdot \hat{s}_1 m_{\chi }+2 m \left(m v \hat{v}\cdot \hat{s}_1-m_{\chi } \left(\hat{s}_1\cdot \hat{s}_3+\hat{v}_{\chi }\cdot \hat{s}_1 v_{\chi }\right)\right)\right)\right) \\
	2\Re(\Delta M_{\rm VA-T_R}(\hs{1},\hs{3})) &=& -4 m \, m_{\chi } \left(m \Im\left(c_{\rm T_R} d_{\rm T_R}\right) \left(m v \hat{v}\cdot (\hat{s}_1\times \hat{s}_3)-\hat{v}_{\chi }\cdot (\hat{s}_1\times \hat{s}_3) m_{\chi } v_{\chi }\right) \right. \\
	&&\quad \left. + \Re\left(c_{\rm T_R} d_{\rm T_R}\right) \left(m v \hat{v}\cdot \hat{s}_1 m_{\chi }+m \left(m v \hat{v}\cdot \hat{s}_1-m_{\chi } \left(\hat{s}_1\cdot \hat{s}_3+\hat{v}_{\chi }\cdot \hat{s}_1 v_{\chi }\right)\right)\right)\right) 
	\end{eqnarray*}
	\noindent
	$\hs{1}\neq0,\;\hs{2}\neq0,\;\hs{3}=0$ case, terms to the order O$(v,v_{\chi})$:
	\begin{eqnarray*}
	\Delta M_{\rm VA}({\hat s}_1,{\hat s}_2) &=& -4 m^2 m_{\chi }^2 \left(\hat{s}_1\cdot \hat{s}_2-\hat{v}_{\chi }\cdot \hat{s}_1 v_{\chi }\right) \\
	\Delta M_{\rm SR}(\hs{1},\hs{2}) &=& 0 \\
	\Delta M_{\rm T_R}(\hs{1},\hs{2}) &=& 16 m \left| c_{\rm T_R}\right| {}^2 \left| d_{\rm T_R}\right| {}^2 m_{\chi } \left(m v \hat{v}\cdot \hat{s}_1 m_{\chi }+2 m \left(m v \hat{v}\cdot \hat{s}_1-m_{\chi } \left(\hat{s}_1\cdot \hat{s}_2+2 \hat{v}_{\chi }\cdot \hat{s}_1 v_{\chi }\right)\right)\right) \\
	2\Re(\Delta M_{\rm VA-S_R}(\hs{1},\hs{2})) &=& -4 m \, m_{\chi } \left(m v \hat{v}\cdot (\hat{s}_1\times \hat{s}_2) \Im\left(c_{\rm S_R} d_{\rm S_R}\right) \left(2 m+m_{\chi }\right)+\Re\left(c_{\rm S_R} d_{\rm S_R}\right) \left(m v \hat{v}\cdot \hat{s}_1 m_{\chi } \right.\right. \\
	&&\quad \left.\left. + 2 m \left(m v \hat{v}\cdot \hat{s}_1+\hat{s}_1\cdot \hat{s}_2 m_{\chi }-\hat{v}_{\chi }\cdot \hat{s}_1 m_{\chi } v_{\chi }\right)\right)\right)\\
	2\Re(\Delta M_{\rm VA-T_R}(\hs{1},\hs{2})) &=& -8 m \, m_{\chi } \left(\Im\left(c_{\rm T_R} d_{\rm T_R}\right) \left(m v \hat{v}\cdot (\hat{s}_1\times \hat{s}_2) \left(m+m_{\chi }\right)-4 m \hat{v}_{\chi }\cdot (\hat{s}_1\times \hat{s}_2) m_{\chi } v_{\chi }\right) \right. \\
	&&\quad \left. + \Re\left(c_{\rm T_R} d_{\rm T_R}\right) \left(m v \hat{v}\cdot \hat{s}_1 m_{\chi }+m \left(m v \hat{v}\cdot \hat{s}_1-m_{\chi } \left(3 \hat{s}_1\cdot \hat{s}_2+\hat{v}_{\chi }\cdot \hat{s}_1 v_{\chi }\right)\right)\right)\right) 
	\end{eqnarray*}
	\noindent
	$\hs{1}\neq0,\;\hs{2}\neq0,\;\hs{3}\neq0$ case, terms to the order O$(v,v_{\chi})$:
	\begin{eqnarray*}
	\Delta M_{\rm VA}(\hs{1},\hs{2},\hs{3}) &=& -2 m^2 \hat{s}_1\cdot \hat{s}_2 m_{\chi }^2 \\
	\Delta M_{\rm S_R}(\hs{1},\hs{2},\hs{3}) &=& 8 m^2 m_{\chi }^2\left| c_{\rm S_R}\right| {}^2 \left| d_{\rm S_R}\right| {}^2 \hat{s}_1\cdot \hat{s}_3 \\
	\Delta M_{\rm T_R}(\hs{1},\hs{2},\hs{3}) &=& -8 m^2 \left| c_{\rm T_R}\right| {}^2 \left| d_{\rm T_R}\right| {}^2 \left(2 \hat{s}_1\cdot \hat{s}_2+\hat{s}_1\cdot \hat{s}_3\right) m_{\chi }^2 \\
	2\Re(\Delta M_{\rm VA-S_R}(\hs{1},\hs{2},\hs{3})) &=& 4 m^2 \left(\hat{s}_1\cdot (\hat{s}_2\times \hat{s}_3) \Im\left(c_{\rm S_R} d_{\rm S_R}\right)-\left(\hat{s}_1\cdot \hat{s}_2-\hat{s}_1\cdot \hat{s}_3\right) \Re\left(c_{\rm S_R} d_{\rm S_R}\right)\right) m_{\chi }^2 \\
	2\Re(\Delta M_{\rm VA-T_R}(\hs{1},\hs{2},\hs{3})) &=& 4 m^2 \left(\hat{s}_1\cdot (\hat{s}_2\times \hat{s}_3) \Im\left(c_{\rm T_R} d_{\rm T_R}\right)+\left(3 \hat{s}_1\cdot \hat{s}_2+\hat{s}_1\cdot \hat{s}_3\right) \Re\left(c_{\rm T_R} d_{\rm T_R}\right)\right) m_{\chi }^2
\end{eqnarray*}

%
% challange, tension, anomaly, inconsistency, instabilty
% problem in DM physics and DM community
% but, although, however
% benefit to reader
%

%\section*{References}

%\bibliographystyle{plain}


\begin{thebibliography}{99}
	%\bibliography{sbbibfile}
	\bibitem{SM} S. L. Glashow, Partial-symmetries of weak interactions,  Nucl. Phys.   22 (1961) 579.
	\bibitem{SM1} S. Weinberg, A Model of Leptons,  Phys. Rev. Lett.   19   (1967) 1264.
	\bibitem{SM2}  A. Salam, in   Elementary Particle Theory (Almquist and Wiksells, Stockholm, 1969).
	\bibitem{SM3} R. P. Feynman, M. Gell-Mann, Theory of the Fermi Interaction, Phys. Rev.  109 (1958)   193.
	\bibitem{SM4} E. C. G. Sudarshan, R. E. Marshak, Chirality Invariance and the Universal Fermi Interaction, Phys. Rev.  109  (1958)  1860.
	\bibitem{barion} A. Riotto, M. Trodden, Recent Progress in Baryogenesis, Annu. Rev. Nucl. Part. Sci.  49 (1999)  35. 
	\bibitem{Kobayashi} M. Kobayashi, T. Maskawa, CP-Violation in the Renormalizable Theory of Weak Interaction, Prog. Theor. Phys.  49 (1973) 652.
	\bibitem{DM1}
	G. Bertone, D. Hooper, J. Silk, Particle dark matter: Evidence, candidates and constraints,
	Phys. Rept. 405 (2005) 279 [hep-ph/0404175].
	\bibitem{DM2} G. Jungman, M. Kamionkowski, K. Griest, Supersymmetric dark matter, Phys. Rept. 267 (1996) 195 [hep-ph/9506380].
	\bibitem{DM3} L. Bergstrom, Non-Baryonic Dark Matter - Observational Evidence and Detection Methods, Rept. Prog. Phys. 63 (2000) 793 [hep-ph/0002126].
	\bibitem{DM4} G. D’Amico , M. Kamionkowski, K. Sigurdson, Dark Matter Astrophysics, 
	 [https://doi.org/10.48550/arXiv.0907.1912].
	\bibitem{DM5} T. Appelquist, H.-C. Cheng, B.A. Dobrescu, 
	Bounds on universal extra dimensions, Phys. Rev. D 64 (2001) 035002 [hep-ph/0012100]. 
	\bibitem{DM6} H.-C. Cheng,
	J.L. Feng, K.T. Matchev, Kaluza-Klein Dark Matter,  Phys. Rev. Lett. 89 (2002) 211301 [hep-ph/0207125]. 
	\bibitem{DM7} G. Servant, T.M.P. Tait, Elastic Scattering and Direct Detection of Kaluza-Klein Dark Matter,  New J.
	Phys. 4 (2002) 99 [hep-ph/0209262].
	\bibitem{DM8} D. Hooper, S. Profumo, Dark Matter and Collider Phenomenology of Universal Extra Dimensions,  Phys. Rept. 453 (2007) 29 
	[hep-ph/0701197]. 
	\bibitem{DM9} G.D. Starkman, D.N. Spergel, Proposed New Technique for Detecting Supersymmetric Dark Matter, Phys. Rev. Lett. 74 (1995) 2623.
	\bibitem{DM10} R. Essig, J. Mardon, T. Volansky,  Direct Detection of Sub-GeV Dark Matter,  	Phys. Rev. D 85 (2012) 076007 [hep-ph/
	1108.5383].
	\bibitem{DM11} P.W. Graham, D.E. Kaplan, S. Rajendran, M.T. Walters, Semiconductor Probes of Light Dark Matter,  Phys. Dark Univ. 1 (2012) 32 [hep-ph/1203.2531].
	\bibitem{DM12} M.W. Goodman, E. Witten, Detectability of certain dark-matter candidates,   Phys. Rev. D 31 (1985) 3059. 
	\bibitem{DM13} I. Wasserman, Possibility of detecting heavy neutral fermions in the Galaxy,  Phys. Rev. D 33 (1986) 2071.
	\bibitem{DM14} D.N. Spergel, Motion of the Earth and the detection of weakly interacting massive particles, Phys. Rev. D 37 (1988) 1353. 
	\bibitem{DM15} P. Gondolo, Recoil momentum spectrum in directional dark matter detector, Phys. Rev. D 66 (2002) 103513 [hep-ph/0209110].
	\bibitem{DM16} M. Lisanti, J.G. Wacker,  Disentangling Dark Matter Dynamics with Directional Detection, Phys. Rev. D 81 (2010) 096005 [hep-ph/0911.1997].
	\bibitem{NuB1} R. Harnik, J. Kopp, P.A.N. Machado, Exploring nu Signals in Dark Matter Detectors,JCAP 07 (2012) 026 [arXiv:1202.6073].
	\bibitem{NuB2} M. Pospelov, J. Pradler, Elastic scattering signals of solar neutrinos with enhanced baryonic currents, Phys. Rev. D 85 (2012) 113016 [arXiv:1203.0545].
	\bibitem{NuB3} D.G. Cerdeño et al., Physics from solar neutrinos in dark matter direct detection experiments, JHEP 05 (2016) 118 [Erratum ibid. 09 (2016) 048] [arXiv:1604.01025].
	\bibitem{NuB4} J.B. Dent, B. Dutta, J.L. Newstead, L.E. Strigari, Dark matter, light mediators and the neutrino floor, Phys. Rev. D 95 (2017) 051701 [arXiv:1607.01468].
	\bibitem{NuB5} J.B. Dent, B. Dutta, J.L. Newstead, L.E. Strigari, Effective field theory treatment of the neutrino background in direct dark matter detection experiments, Phys. Rev. D 93 (2016) 075018 [arXiv:1602.05300].
	\bibitem{NuB6} P. Grothaus, M. Fairbairn, J. Monroe, Directional dark matter detection beyond the neutrino bound, Phys. Rev. D 90 (2014) 055018 [arXiv:1406.5047]. 
	\bibitem{NuB7} E.  Bertuzzo, F. F. Deppisch, S. Kulkarni, Y. F. Perez Gonzalez, R.  Zukanovich Funchal, Dark matter and exotic neutrino interactions in direct
	detection searches, JHEP 04 (2017) 073. 
	\bibitem{PET1} P. Minkowski, M. Passera, Elastic Scattering of Neutrinos off Polarized Electrons,  Phys. Lett.  B  541 (2002)  151.
	\bibitem{PET2} T. I. Rashba, V. B. Semikoz, Neutrino scattering on polarized electron target as a test of neutrino magnetic moment,  Phys. Lett.  B  479 (2000)  218. 
	\bibitem{PET3} J. Bernabeu J.  Papavassiliou, M. Passera, Dynamical zero in $\overline{\nu}_e e^{-}$
	scattering and the neutrino magnetic moment Phys. Lett.  B  613 (2005) 162. 
	\bibitem{PET4} S. Ciechanowicz,  W Sobków, M Misiaszek,  Scattering of neutrinos on a polarized electron target as a test for new physics beyond the standard model,  Phys. Rev.  D  71  (2005)  093006.
	\bibitem{PET5} W. Sobk\'ow, A. B\l{}aut, On possibility of time reversal symmetry violation in neutrino elastic scattering on polarized electron target, Eur. Phys. J. C   78  (2018)  197. 
	\bibitem{PET6}  A. B\l{}aut, W. Sobk\'ow, Neutrino elastic scattering on polarized electrons  as a tool for probing the neutrino nature, Eur. Phys. J. C   80  (2020) 261.
	\bibitem{PDM1} C.-T. Chiang, M. Kamionkowski, G. Z. Krnjaic, Dark Matter Detection with Polarized Detectors, Phys. Dark Univ. 1 (2012) 109 [1202.1807].
	\bibitem{PDM0} R. Catena, K. Fridell, V Zema, Direct detection of fermionic and vector dark matter with polarised targets, JCAP  11 (2018) 018 [1810.01515].
	\bibitem{PDM2} T. Franarin, M. Fairbairn, Reducing the solar neutrino background in dark matter searches using polarized helium-3, Phys. Rev. D94 (2016) 053004 [1605.08727].
	\bibitem{Misiaszek} M. Misiaszek, S Ciechanowicz, W Sobków, The polarized electron target as a new solar-neutrino detector,  Nucl. Phys.  B  734 (2006)  203. 
	\bibitem{INFN} B. Babussinov et al., An active electron polarized scintillating GSO target for neutrino physics,  Nucl. Instrum. and Meth. A  694 (2012) 335.
	\bibitem{Gass1} M. A. Bouchiat, T. R. Carver, C. M. Varnum, Nuclear Polarization in $He^{3}$ Gas Induced by Optical Pumping and Dipolar Exchange,  Phys. Rev. Lett. 5 (1960) 373.  
	\bibitem{Gass2} T. G. Walker, W. Happer, Spin-exchange optical pumping of noble-gas nuclei, Rev. Mod. Phys. 69 (1997) 629.   
	\bibitem{PCP} R. Catena, J. Hagel, C. Yaguna, Probing P- and CP-violation in dark matter
	interactions, JCAP 05 (2021) 016 [2007.01262]. 
	
	
\end{thebibliography}
\end{document}